\documentclass[fleqn,10pt]{wlscirep}

\usepackage[utf8]{inputenc}
\usepackage[T1]{fontenc}
\usepackage{graphicx}
\usepackage{multirow}
\usepackage{amsmath,amssymb,amsfonts}
\usepackage{amsthm}
\usepackage{mathrsfs}
\usepackage[title]{appendix}
\usepackage{xcolor}
\usepackage{textcomp}
\usepackage{manyfoot}
\usepackage{booktabs}
\usepackage{algorithm}
\usepackage{algorithmicx}
\usepackage{algpseudocode}
\usepackage{listings}
\usepackage{float}
\usepackage{tabularx}
\usepackage{gensymb}
\usepackage{booktabs}
\usepackage{titlesec}
\usepackage{lmodern}
\usepackage{anyfontsize}
\usepackage{lineno}
\usepackage{soul}
\setstcolor{red}

\titleformat{\paragraph}[runin]{\normalfont\normalsize\bfseries}{\thesubsubsection}{1em}{}
\newcolumntype{Y}{>{\raggedleft\arraybackslash}X}

\title{\textbf{Towards deep-learning based detection and quantification of intestinal metaplasia on digitized gastric biopsies: a multi-expert comparative study}}

\author[1]{Fabian Cano}
\author[1]{Mauricio Caviedes}
\author[1]{Andres Siabatto}
\author[1]{Jesus Villarreal}
\author[1]{Jose Quijano}
\author[5]{Álvaro Bedoya-Urresta}
\author[6]{Marino Coral Bedoya}
\author[5]{Yomaira Yepez Caicedo}
\author[3]{Angel Cruz-Roa}
\author[2]{Fabio A. González}
\author[4]{Satish E. Viswanath}
\author[1,*]{Eduardo Romero}

\affil[1]{Computer Imaging and Medical Applications Laboratory - Cim@lab, Universidad Nacional de Colombia, Bogotá D.C., Colombia}

\affil[2]{Machine Learning, Perception and Discovery Lab - Mindlab, Universidad Nacional de Colombia, Bogotá D.C., Colombia}

\affil[3]{GITECX Research Group \& Automatic Data-driven Analytics Laboratory - AdaLab, Universidad de los Llanos, Meta, Colombia}

\affil[4]{Department of Biomedical Engineering, Case Western Reserve University, Cleveland, United States}

\affil[5]{Centro de Investigación de Enfermedades digestivas y nutricionales - CIEDYN, Hospital Universitario Departamental de Nariño, Pasto, Colombia}

\affil[6]{Faculty of Health Sciences, Universidad Cooperativa de Colombia, Pasto, Colombia}

\affil[*]{Corresponding authors: edromero@unal.edu.co}

\begin{abstract}
    Current gastric cancer (GCa) risk systems are prone to errors since they evaluate a visual estimation of intestinal metaplasia percentages in histopathology images of gastric mucosa to assign a risk. This study presents an automated method to detect and quantify intestinal metaplasia using deep convolutional neural networks as well as a comparative analysis with visual estimations of three experienced pathologists. Gastric samples were collected from two different cohorts: 149 asymptomatic volunteers from a region with a high prevalence of GCa in Colombia and 56 patients from a tertiary hospital. Deep learning models were trained to classify intestinal metaplasia, and predictions were used to estimate a percentage of intestinal metaplasia and to assign an adapted OLGIM stage. Atrophy was not assessed because of the limited reproducibility among pathologists. Results were compared with independent blinded metaplastic assessments performed by three graduated pathologists. The best-performing deep learning architecture classified intestinal metaplasia with F1-Score of $0.80 \pm 0.01$ and AUC of $0.91 \pm 0.01$. Among pathologists, inter-observer agreement by a Fleiss's Kappa score ranged from $0.20$ to $0.48$. In comparison, agreement between the pathologists and the best-performing model ranged from $0.12$ to $0.35$. Deep learning models show potential to reliably detect and quantify the percentage of intestinal metaplasia, achieving high classification performance. In practice, visual estimation is still the only available method, yet it is marked by considerable inter-observer variability. Deep learning models provide consistent estimates that could help reduce this subjectivity in risk stratification.
\end{abstract}


\begin{document}
\maketitle

\section*{Introduction}
\label{sec:introduction}

Gastric cancer (GCa) was the fifth most commonly diagnosed cancer and the fifth leading cause of cancer-related death worldwide in 2022 \cite{Globocan2022}. Approximately 90\% of GCa cases were diagnosed as adenocarcinomas, from which the most common type was intestinal adenocarcinoma~\cite{Ajani2017}. The high mortality rate associated with GCa is closely related to asymptomatic progression, however, if diagnosed at early stages, patient survival improves considerably~\cite{Tan2015}. The way in which this cancer develops and progresses is not completely clear, probably a combination of genetic factors associated with bacterial aggressiveness, the \textit{Helicobacter pylori}, and environmental or lifestyle factors~\cite{Pucułek2018}. Given the lack of effective strategies to cure GCa, early diagnosis remains the most promising strategy to reduce both incidence and mortality rates~\cite{Cai2019}.

Currently, the screening protocol is guided by the Sydney System, introduced in 1991~\cite{Misiewicz1991} and updated in 1994 as the Updated Sydney System (USS)~\cite{Dixon1996}. This system recommends histological evaluation of at least five gastric biopsies obtained from both the antrum and corpus, three from lesser curvature and two from greater curvature. This estimator has been popularized in practice because the incisura angularis has been considered an area with higher risk and therefore has been included as an additional antrum biopsy~\cite{Rugge2007}, and some studies have observed premalignant lesions in this region~\cite{Crafa2018}. Overall, \textit{Helicobacter pylori}, mononuclear cells, loss of glands (atrophy), intestinal metaplasia, inflammation or dysplasia can be found in any of the five collected biopsies. Particular attention has been paid to intestinal metaplasia which has been scored in different categories: 0 (absence), 1 (mild), 2 (moderate), and 3 (severe), a combination of the extent and topographic distribution of microscopic changes~\cite{Correa2013}. This type of lesion is considered a gastric adaptation to chronic infection with \textit{Helicobacter pylori}~\cite{Koulis2019} and a pivotal event in GCa progression, described as a \textit{``point of no return"}~\cite{Tjandra2023}. The final step to establish a progression risk to GCa is to assign a stage~\cite{Marcos2020}. Protocols developed to assign this risk are basically the Operative Link for Gastritis Assessment (OLGA) system~\cite{Rugge2005} and the Operative Link on Gastric Intestinal Metaplasia (OLGIM) system~\cite{Capelle2010} (specifically designed to stage the extent and distribution of intestinal metaplasia). Both systems classify patients from stage 0 (lowest risk) to stage IV (highest risk).

Recently, some concerns have been raised about the uncertainty and clinical implications of current gastric risk stratification~\cite{Fang2024,Zhao2022} after the original USS version was introduced. Several publications have reported variability among general pathologists when assigning OLGIM stages~\cite{Fang2024,Yue2018,Isajevs2014} and they have documented that these systems may underestimate or overestimate cancer risk~\cite{Yue2018,Molaei2016,Crafa2018}. Although some studies suggest that OLGIM may stratify GCa risk~\cite{Capelle2010,Rugge2010}, there is currently no robust quantitative evidence to support this statement~\cite{Mansour2022,Yue2018}. Therefore, a more precise and reproducible method to quantify these premalignant lesions remains an unmet need~\cite{Fang2024}. In this context, automatic approaches are particularly appealing as they address many of the limitations associated with visual assessment. Deep learning, in particular, has demonstrated high accuracy for detecting and quantifying patterns in medical images~\cite{Fang2024}, with outstanding results reported in lung cancer~\cite{Chen2021,Coudray2018,Kanavati2020}, breast cancer~\cite{Cruz2018,LiuM2022,Wetstein2022}, prostate cancer~\cite{Nir2018,Li2020,Tolkach2020}, and gastric cancer~\cite{Wang2021,Huang2021,Sharma2017}. Specifically in gastric cancer, deep learning models have shown improved diagnostic accuracy and reproducibility in histopathology images, while significantly reducing both inter- and intra-observer variability~\cite{Veldhuizen2023}.

The current OLGIM risk assessment system, which routinely guides in treatment of gastric cancers~\cite{White2022,Tjandra2023}, assigns a risk score by heuristically combining visual estimations of intestinal metaplasia from five different gastric biopsy sites. In this context, intestinal metaplasia remains a critical premalignant GCa stage that requires accurate and reproducible quantification~\cite{Fang2024}. Interestingly, some studies have reported cases classified as OLGIM low-risk who were later diagnosed with gastric cancer~\cite{Pimentel2019,Mansour2022}, while others identified as high-risk never developed cancer and, in some cases, reversed the lesion~\cite{Hwang2018,Aumpan2021}. In summary, these investigations indicate that subjectively estimated GCa risk is prone to errors and dependent on expert evaluation~\cite{Tjandra2023}. 

In this study, an automatic deep learning approach is proposed to accurately detect and quantify intestinal metaplasia in hematoxylin and eosin (H\&E)-stained images. A main contribution consists in applying and adapting several state-of-the-art neural networks to compute the percentage of intestinal metaplasia, and their performance was compared with manual estimations made by three pathologists. Evaluation with two complementary cohorts, one from a high-risk population of Colombian volunteers and another from a tertiary hospital, allowed assessment of classification accuracy and robustness across different clinical contexts.

\section*{Materials and Methods}
\label{sec:methods}

\begin{figure}[H]
    \centering
    \includegraphics[width=1\linewidth]{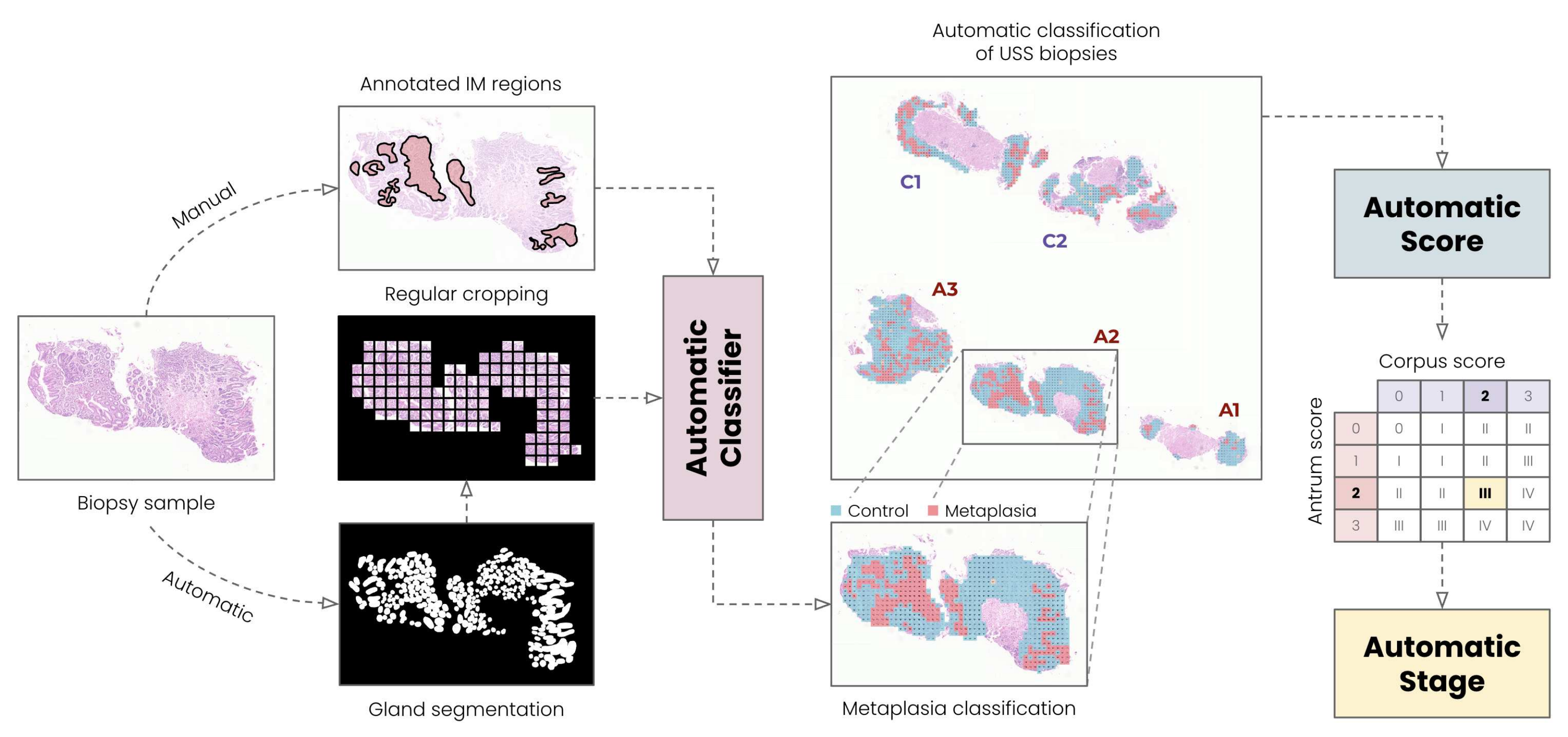}
    \caption{Workflow of the automatic intestinal metaplasia classification with deep learning models and automatic scoring using the proportion of intestinal metaplasia in both antrum and corpus biopsies.}
    \label{fig:methodology}
\end{figure}

\subsection*{Data Description}
\label{subsec:dataset}

Gastric biopsy samples from two independent cohorts were digitized. The first cohort consists of 149 asymptomatic volunteers who were selected from an independent study provided by the \textit{CIEDYN} foundation, a partner of the \textit{Urkunina5000} project~\cite{Urkunina5000}. This study performed an endoscopic and pathological characterization of gastric disease in 5.000 volunteers from 55 small villages in \textit{Nariño}, Colombia, a region long observed as being GCa prevalent for the past two decades. Asymptomatic volunteers were men and women aged $30-70$ years, who had lived in these areas for at least the last 10 years and had never undergone an endoscopy procedure, nor been diagnosed with GCa, any other type of cancer, or any known gastric pathology. The second cohort included 56 patients from the \textit{Hospital Universitario Nacional de Colombia} included in a larger study which was approved by the ethics committee (Act No. 007, Apr 29 2022). Patients were men and women who present symptoms suggestive of gastritis or other forms of gastric discomfort. However, none of them had a previous GCa diagnosis or any other type of lesion.

\renewcommand{\arraystretch}{1.2}
\begin{table}[ht]
    \caption{Demographic characteristics and OLGIM stage distribution for both data cohorts. The external test set corresponds exclusively to the second cohort.}
    \label{tab:dataset}
    \begin{tabularx}{\textwidth}{@{} l Y Y Y Y Y @{}}
        \toprule
        \textbf{Characteristics} & \textbf{All cases} (N=205) & \textbf{Training} (N=73) & \textbf{Validation} (N=32) & \textbf{Internal Test} (N=44) & \textbf{External Test} (N=56) \\
        \midrule
        Age (years) & 53.68 $\pm$ 10.39 & 52.68 $\pm$ 10.68 & 56.62 $\pm$ 10.51 & 53.20 $\pm$ 9.61 & 64.73 $\pm$ 13.63 \\
        Female (\%) & 79 (53.02\%) & 44 (60.27\%) & 13 (40.62\%) & 22 (50.00\%) & 38 (67.86\%) \\
        \textbf{OLGIM} &  &  &  &  & \\
        \hspace{0.5cm} 0 & 74 (36.10\%) & 16 (21.92\%) & 7 (21.88\%) & 10 (22.73\%) & 41 (73.21\%) \\
        \hspace{0.5cm} 1 & 74 (36.10\%) & 31 (42.47\%) & 14 (43.75\%) & 21 (47.73\%) & 8 (14.29\%) \\
        \hspace{0.5cm} 2 & 34 (16.59\%) & 14 (19.18\%) & 7 (21.88\%) & 8 (18.18\%) & 5 (8.93\%) \\
        \hspace{0.5cm} 3 & 14 (6.83\%) & 9 (12.33\%) & 2 (6.25\%) & 3 (6.82\%) & 0 (0.00\%) \\
        \hspace{0.5cm} 4 & 9 (4.39\%) & 3 (4.11\%) & 2 (6.25\%) & 2 (4.55\%) & 2 (3.57\%) \\
        \bottomrule
    \end{tabularx}
\end{table}

\subsubsection*{Biopsy assessment}
\label{subsubsec:biopsy}

\begin{figure}[H]
    \centering
    \includegraphics[width=1\linewidth]{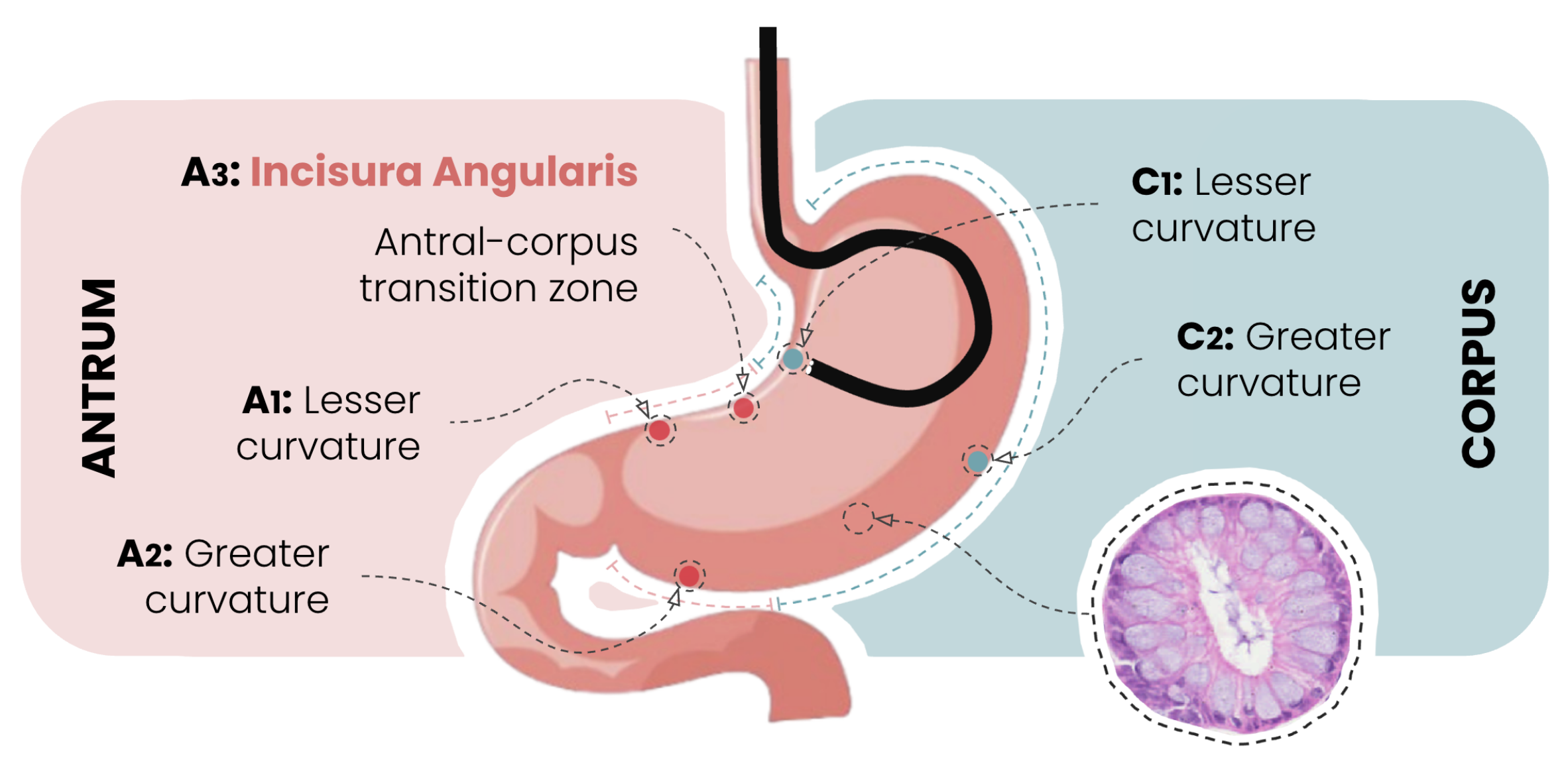}
    \caption{Biopsy places established for GCa detection according to the Updated Sydney System (USS). Antrum: Lesser curvature, Greater curvature and Antral-corpus transition zone (incisura angularis). Corpus: Lesser curvature and Greater curvature.}
    \label{fig:sydneyandolgim}
\end{figure}

Five gastric biopsies were obtained according to the USS (Fig. \ref{fig:sydneyandolgim}) in both cohorts. Specifically, two biopsies were taken from the antrum ($A_1$, $A_2$), other two biopsies from the corpus ($C_1$, $C_2$), and a final biopsy from the incisura angularis ($A_3$) (antral-corpus transition zone). All biopsies were fixed in formalin and embedded in paraffin blocks and stained with H\&E

All cases were digitized using a whole-slide image (WSI) scanner (MoticEasyScan Pro), with a $\times 40$ objective, corresponding to a spatial resolution of $0.255 \: \mu m$ per pixel (MPP). Each case comprised five biopsies (one per anatomical region of the stomach), resulting in a total of 205 WSI, i.e., 149 from the first cohort and 56 from the second. The first cohort was used to train and evaluate the deep learning models including 73 cases for training, 32 cases for validation and 44 cases for testing. The second cohort was used exclusively as an external testing set.

Each case was independently assessed by three experienced pathologists following the OLGIM system. This system assumes that variations regarding extent and topographical distribution of intestinal metaplasia reflect distinct clinic-pathological scenarios, each corresponding to a specific GCa risk. Pathologists performed blinded evaluations to estimate the proportion of intestinal metaplasia at five standardized biopsy sites: three from the mucosecreting area (two from the antrum and one from the incisura angularis) and two from the oxyntic mucosa (lesser and greater curvature of the corpus). After intestinal metaplasia was detected, a score was assigned as the estimated percentage of glands with goblet cells, ideally assessed in full-thickness (perpendicular) mucosa sections. Each biopsy was independently scored following a four-tier scale: absence of metaplasia = 0\% (score 0), mild metaplasia = 1 – 30\% (score 1), moderate metaplasia = 31 – 60\% (score 2), and severe metaplasia = > 60\% (score 3). The scores for the antral biopsies ($A_1$, $A_2$, $A_3$) and the corpus biopsies ($C_1$, $C_2$) were averaged separately to obtain a composite score for each region. Finally, risk was determined by combining mean scores of the antrum and corpus, as illustrated in Fig. \ref{fig:sydneyandolgim}.

\subsubsection*{Manual annotations of intestinal metaplasia}
\label{subsub:expertannotations}

All cases were manually annotated by the most experienced pathologist, delineating those regions affected by intestinal metaplasia. Overall, these regions contain not only glands with intestinal metaplasia but also a non-negligible amount of stroma and surrounding tissues. The pathologist manually annotated intestinal metaplasia regions using a custom web software developed as part of the \textit{Program for the Early Detection of Premalignant Lesions and Gastric Cancer in urban, rural and dispersed areas in the Department of Nariño}, equipped with basic tools to annotate and visualize WSI. Annotations were saved according to the Web Annotation Data Model standard \cite{Sanderson2017}.

Additionally, all cases were evaluated by the three pathologists, who independently estimated the percentage of intestinal metaplasia in both antrum and corpus biopsies using the previously described four-tier scale. Based on these estimations, each pathologist assigned an OLGIM stage per case, derived from the previously mentioned heuristic combination. This staging reflects the presumed risk of gastric cancer progression associated with the distribution and severity of intestinal metaplasia.

\subsection*{Automatic detection of glandular regions}
\label{subsubsec:segmentation}

The initial step to detect and quantify intestinal metaplasia involves automatic segmentation of gastric glands. In this study, gland segmentation was performed applying a U-Net architecture with a ResNet18 backbone~\cite{Sirinukunwattana2017}, originally trained to segment colorectal glands. This choice was motivated by the fact that colorectal and gastric glands exhibit remarkable morphological similarities~\cite{Jass2007}, particularly by the presence of goblet cells associated with intestinal metaplasia, which by definition corresponds to structural changes resembling those observed in colorectal tissue~\cite{Shah2020}.

The model was pre-trained using a dataset comprising 165 fields of view extracted from 16 H\&E-stained whole slide images (WSI) from the GlaS database (Gland Segmentation in Colon Histology Images Challenge)~\cite{Sirinukunwattana2017}. Since gastric and colorectal glands are similar in appearance, this architecture processed the 205 WSI from both gastric cohorts. To adapt the model to the gastric domain, the most experienced pathologist manually annotated 2.434 glands from 15 randomly selected cases, including both metaplastic and non-metaplastic glands. This annotated dataset was used to fine-tune the U-Net architecture, achieving a Dice Score of $0.77 \pm 0.22$ with a test subset of 730 glands, that is, glands not included in the training phase.

Since the objective of this study is not only to segment glands but also to detect and quantify intestinal metaplasia, the segmented glands were used to define glandular regions. These regions were defined as bounding boxes surrounding groups of glands, within which a grid of fields of view is drawn. 

\subsection*{Labeling automatically detected fields of view}
\label{subsubsec:patchextraction}

Intestinal metaplasia regions, manually annotated by the most experienced pathologist, were superimposed onto the previously described grid of fields of view. Fields overlapping the annotated regions were labeled as intestinal metaplasia. Each of these fields of view corresponds to square patches of $256 \times 256$ pixels, extracted at $\times40$ magnification. Regions outside the glandular region were excluded from the analysis. 

The 149 cases from the first cohort were divided into training, validation, and internal testing sets, while the 56 cases from the second cohort were exclusively used as an external test set. In total, $476.351$ fields of view were extracted, being $136.682$ for training, $59.629$ for validation, $89.945$ for internal testing, and $190.095$ for external testing.

\subsection*{Automatic classification of intestinal metaplasia}
\label{subsubsec:classification}

\subsubsection*{Deep learning models} Fields of view labeled as metaplasia or control classes were used to train and evaluate a set of state-of-the-art deep learning models. Specifically, four state-of-the-art convolutional neural networks pre-trained with the ImageNet dataset \cite{Deng2009}, ResNet50 \cite{Resnet2016}, DenseNet121 \cite{DenseNet121} and ConvNeXtTiny \cite{Liu2022}, were adapted to classify microscopic gastric fields. ResNet50 was selected based on its remarkable performance in intestinal metaplasia classification, with predictions closely matching the assessments of experienced pathologists \cite{Caviedes2023}. DenseNet121 was used as the basis to construct a specialized histopathology framework \cite{KimiaNet}. ConvNeXtTiny has demonstrated superior performance at identifying anatomical regions of the stomach in endoscopic images \cite{Bravo2023}. Furthermore, a foundational model, UNI2-h~\cite{Chen2024}, pre-trained with TCGA~\cite{Tomczak2015}, a large-scale histopathology image dataset, was used as a feature extractor for the classification of intestinal metaplasia.

\subsubsection*{Warm-up and fine-tuning} A multi-layer perceptron (MLP) was trained on top of each model to perform the classification of intestinal metaplasia. All models followed a two-stage training scheme consisting of a warm-up and fine-tuning phase. During the warm-up phase, the backbone of each model was frozen to preserve visual features learned from large datasets, such as ImageNet or TCGA. In this phase, only the MLP weights were updated. Subsequently, in the fine-tuning phase, selected layers of the pre-trained backbone were progressively unfrozen, enabling joint optimization of both the backbone and the classification head. Fine-tuning was performed with a lower learning rate to minimize overfitting and preserve the most relevant features of the original pretraining.

\subsubsection*{Hyper-parameter optimization} Each model was trained, validated and tested using the data partition presented in section \textit{Biopsy assessment}, along with the corresponding fields of view described in section \textit{Labeling automatically detected fields of view}. The architecture of each model integrated a MLP consisting of a sequential block composed of a dropout layer, a fully-connected layer, and a batch normalization layer, followed by an output layer with a \textit{softmax} activation function. Model training followed a two-phase scheme (warm-up and fine-tuning), each consisting of thirty epochs. A hyper-parameter optimization process was performed to determine the optimal learning rate and dropout values, based in three independent experimental runs. During the warm-up phase, only the MLP was trained, while the backbone of the pre-trained model remained frozen. In the subsequent fine-tuning phase, 80\% of the pre-trained model layers remained frozen, while the remaining 20\% (the deepest layers) were unfrozen and jointly optimized with the MLP to refine task-specific representations. 

\subsubsection*{Data augmentation strategy} A significant class imbalance was noted for the dataset, with a ratio 9:1 for non-metaplasia vs metaplasia patches. To address this imbalance, an online data augmentation strategy was implemented during training. This strategy involved applying symmetrical rotations of $90^\circ$ increments along the horizontal axis to artificially increase sample diversity and mitigate overfitting. Furthermore, class weights were adjusted before training, assigning greater importance to the minority class (\textit{metaplasia}) in the loss function to compensate for the skewed class distribution.

\subsection*{Automatic quantification and scoring of intestinal metaplasia}
\label{subsubsec:estimation}

Automatic quantification of intestinal metaplasia was performed independently by the four deep learning architectures at the level of the pre-established fields of view. Each field of view was set to a probability value of class membership (i.e., \textit{metaplasia} or \textit{control}), and model performance was assessed using standard metrics: Accuracy, Precision, Recall, F1-Score and Area Under the ROC Curve (AUC). 

These regional proportions were then used to determine an intestinal metaplasia score according to the OLGIM system, by averaging the scores from the three antrum biopsies ($A_1$, $A_2$, $A_3$) and the two corpus biopsies ($C_1$, $C_2$). These scores reflect both the extent and topographic distribution of intestinal metaplasia, and were further used to assign the OLGIM stage by intersecting the antrum and corpus scores. Stages 0 - II were classified as low-risk, while stages III - IV were considered high-risk for GCa progression (Fig.~\ref{fig:sydneyandolgim}).

\subsection*{Quantifying variability of manual and automatic estimations}
\label{subsubsec:interexpert}

Manual estimations were performed by three pathologists following OLGIM guidelines. Pathologists assigned scores approximating the extent and topographic distribution of intestinal metaplasia across the antrum, incisura, and corpus

\subsubsection*{Inter-observer variability} The 205 cases were independently evaluated by three pathologists following OLGIM guidelines. Pathologists assigned scores approximating the extent and topographic distribution of intestinal metaplasia across the antrum, incisura, and corpus, and assigned a score from 0 to 3. Based on these scores, an OLGIM stage was assigned: stages 0, I or II indicate the lowest risk, while stages III and IV indicate the highest risk. To assess inter-observer agreement, Fleiss' Kappa coefficient was calculated to measure overall consistency among the three pathologists, and Cohen's Kappa coefficient was used to assess pairwise agreement. Both metrics range from 0 (no agreement) to 1 (perfect agreement) with higher values indicating greater agreement.

\subsubsection*{Comparison between manual and automatic intestinal metaplasia percentages} Automatic scores calculated by the deep learning models were compared against the manual estimations provided by three experienced pathologists in two complementary approaches. 

First, trends and variability of antrum and corpus intestinal metaplasia percentages are plotted to show differences between the three pathologists and four deep learning models. Second, a separate plot illustrates the distribution of the assigned OLGIM stages, allowing a direct comparison of staging variability between manual assessments and automated predictions.

\section*{Results}
\label{sec:results}

\subsection*{Automatic classification performance}
\label{subsec:performance}

Deep learning models were fine-tuned and evaluated in a binary classification task (presence vs. absence of intestinal metaplasia) using both cohorts. Specifically, models were trained, validated and tested using the first cohort of 149 cases, and externally tested with the second independent cohort of 56 cases. The classification results, summarized in Table~\ref{tab:performance}, show mean and standard deviation for both internal and external test sets. These metrics were calculated across three experimental runs during the hyperparameter optimization phase and test sets were left aside from the beginning to ensure unbiased evaluation. Under these conditions, ConvNeXtTiny achieved the highest performance when classifying fields of view, with F1-Score \textbf{0.80} $\pm$ \textbf{0.01}, and AUC \textbf{0.91} $\pm$ \textbf{0.01}, for the internal test set, and F1-Score \textbf{0.73} $\pm$ \textbf{0.01}, and AUC \textbf{0.83} $\pm$ \textbf{0.01} for the external test set. The UNI2-h foundational model, showed the poorest performance for intestinal metaplasia classification with F1-Score \textbf{0.72} $\pm$ \textbf{0.01} and AUC \textbf{0.84} $\pm$ \textbf{0.01}, for the internal test set, and F1-Score \textbf{0.62} $\pm$ \textbf{0.01} and AUC \textbf{0.75} $\pm$ \textbf{0.01} for the external test set.

\setlength{\tabcolsep}{4pt}
\renewcommand{\arraystretch}{1.2}
\begin{table}[ht]
    \caption{Performance for the field-of-view-level classification models with the internal (1st cohort) and external (2nd cohort) testing sets.}
    \label{tab:performance}
    \begin{tabularx}{\textwidth}{@{} l Y Y Y Y Y Y @{}}
        \toprule
        \textbf{Architecture} & \textbf{Precision} & \textbf{Sensitivity} & \textbf{F1-Score} & \textbf{AUC} \\
        \midrule
        \addlinespace
        \textbf{Internal test} &  &  &  &  \\[0.1cm]
        \hspace{0.3cm} ConvNeXtTiny & 0.79 $\pm$ 0.01 & \textbf{0.80 $\pm$ 0.01} & \textbf{0.80 $\pm$ 0.01} & \textbf{0.91 $\pm$ 0.01} \\
        \hspace{0.3cm} ResNet50 & 0.82 $\pm$ 0.01 & 0.77 $\pm$ 0.01 & 0.79 $\pm$ 0.01 & 0.90 $\pm$ 0.01 \\
        \hspace{0.3cm} DenseNet121 & \textbf{0.83 $\pm$ 0.02} & 0.74 $\pm$ 0.03 & 0.77 $\pm$ 0.02 & 0.89 $\pm$ 0.01 \\
        \hspace{0.3cm} UNI2-h & 0.75 $\pm$ 0.02 & 0.70 $\pm$ 0.02 & 0.72 $\pm$ 0.01 & 0.84 $\pm$ 0.01 \\[0.1cm]
        \textbf{External test} &  &  &  & \\[0.1cm]
        \hspace{0.3cm} ConvNeXtTiny & 0.78 $\pm$ 0.01 & \textbf{0.69 $\pm$ 0.01} & \textbf{0.73 $\pm$ 0.01} & \textbf{0.83 $\pm$ 0.01} \\
        \hspace{0.3cm} ResNet50 & 0.80 $\pm$ 0.01 & 0.64 $\pm$ 0.01 & 0.68 $\pm$ 0.01 & 0.79 $\pm$ 0.01 \\
        \hspace{0.3cm} DenseNet121 & 0.81 $\pm$ 0.04 & 0.64 $\pm$ 0.02 & 0.68 $\pm$ 0.01 & 0.82 $\pm$ 0.01 \\
        \hspace{0.3cm} UNI2-h & \textbf{0.81 $\pm$ 0.01} & 0.51 $\pm$ 0.02 & 0.62 $\pm$ 0.01 & 0.75 $\pm$ 0.01 \\
        \bottomrule
    \end{tabularx}
\end{table}

\subsection*{Automatic quantification and scoring of intestinal metaplasia}

\begin{figure}[htp]
    \centering
    \includegraphics[width=0.8\linewidth]{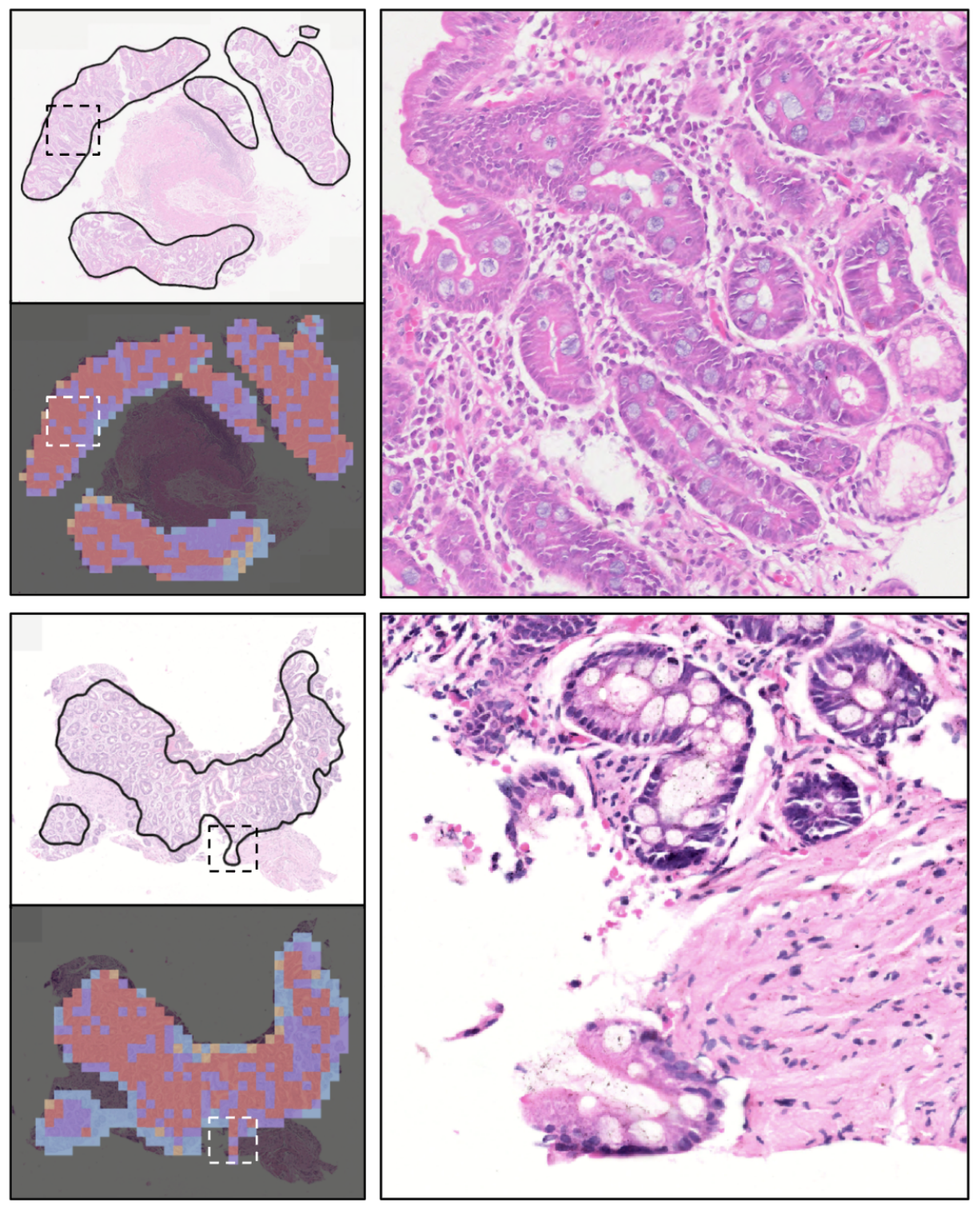}
    \caption{Representative examples comparing expert annotations and model predictions. For each case, the left subfigures display the whole-slide region with the most experienced pathologist’s annotation (top) and the corresponding model prediction map (bottom). Marked boxes indicate areas of interest, which are shown at higher magnification on the right. The enlarged fields illustrate regions containing metaplastic glands, as well as portions of tissue annotated by the expert where not all structures correspond to intestinal metaplasia. Model predictions are color-coded as follows: correctly predicted metaplasia (red), correctly predicted control (blue), control regions incorrectly predicted as metaplasia (orange), and metaplastic regions incorrectly predicted as control (purple).}
    \label{fig:heatmap}
\end{figure}

The fields of view predicted by the best-performing model were superimposed to the corresponding whole-slide images of both cohorts. Predictions were plotted in a color-code map: blue and red, the ones correctly classified, and orange and purple, the misclassified ones (Fig.~\ref{fig:heatmap}). These prediction maps highlight in red areas where the model agrees with the expert-annotated regions. Within these annotated regions, areas in purple correspond to those in which the model disagrees with the expert. Interestingly, the model also identifies metaplastic areas (in orange) that have not previously been annotated. In general, a majority of misclassified fields of view were observed within large regions, which often contain a heterogeneous mix of metaplastic glands, control glands, stroma, and foveolar epithelium.

\subsection*{Quantifying variability of manual and automatic estimations}

\subsubsection*{Results of the inter-observer agreement}
\label{subsubsec:agreement}

\begin{figure}[htp]
    \centering
    \includegraphics[width=1\linewidth]{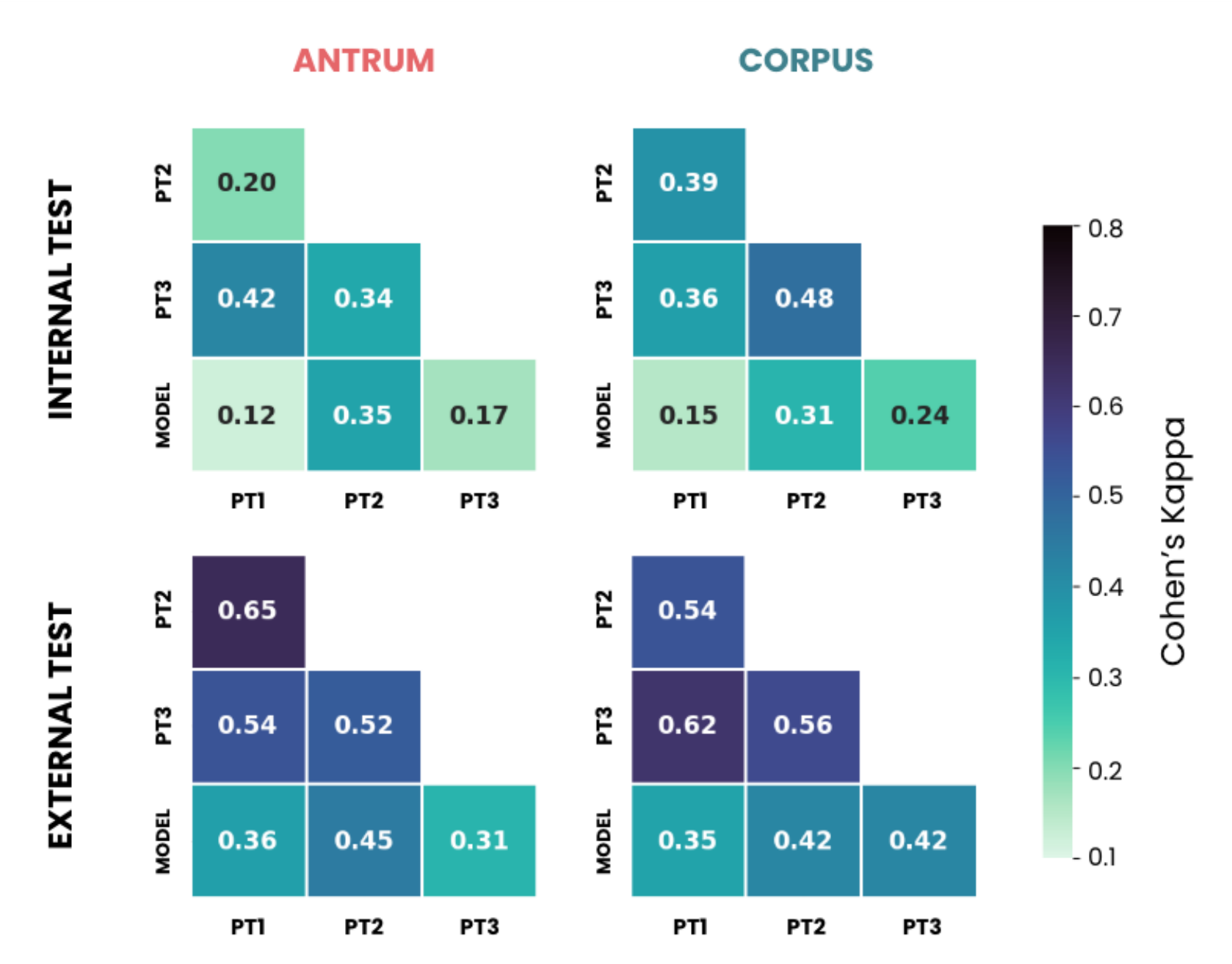}
    \caption{Inter-observer agreement in the assignment of intestinal metaplasia (IM) percentage intervals (10\% bins across the full range) among pathologists and the best-performing deep learning model (ConvNeXtTiny) for the internal (first row) and external (second row) test sets. PT1, PT2 and PT3 correspond to the three pathologists, while MODEL refers to the automated model. Subplots in the first column show agreement for the antrum, and subplots in the second column correspond to the corpus.}
    \label{fig:kappa}
\end{figure}

Inter-observer agreement among pathologists and the best-performing deep learning model is presented in Figure~\ref{fig:kappa}. The Cohen’s Kappa coefficient shows pairwise inter-observer agreement in both the antrum and corpus, for both internal and external test sets. Given that experts provided an accurate estimation of the intestinal metaplasia percentage, manual-automatic comparison was performed at a finer level by binning the entire range into incremental intervals of 10\%, i.e., ten intervals of 10\% instead of the usual 0\%, 30\%, 60\% and more. This uniform partition was used to evaluate agreement between manual and automated estimations. Inter-observer agreement among pathologists was moderate, with Fleiss’ Kappa = $0.31$ for antrum, and $0.41$ for corpus, therefore, pairwise Cohen’s Kappa ranged $0.20$ – $0.48$. Agreement between the model and individual pathologists ranged $0.12$ – $0.35$. Variability of IM percentage estimates was evident, since pathologists tended to assign higher values than the automated model. These results underscore both the visual estimation variability and the relative consistency of the automated approach. In addition, Spearman’s rank-order correlation was computed between the IM percentage estimates of the model and those of each pathologist, showing significant positive associations ($\rho$ = $0.732 \pm 0.011$ for OLGIM stage and $\rho$ = $0.893 \pm 0.022$ for IM percentage), indicating model prediction are associated with pathologists’ estimation

\begin{figure}[htp]
    \centering
    \includegraphics[width=0.8\linewidth]{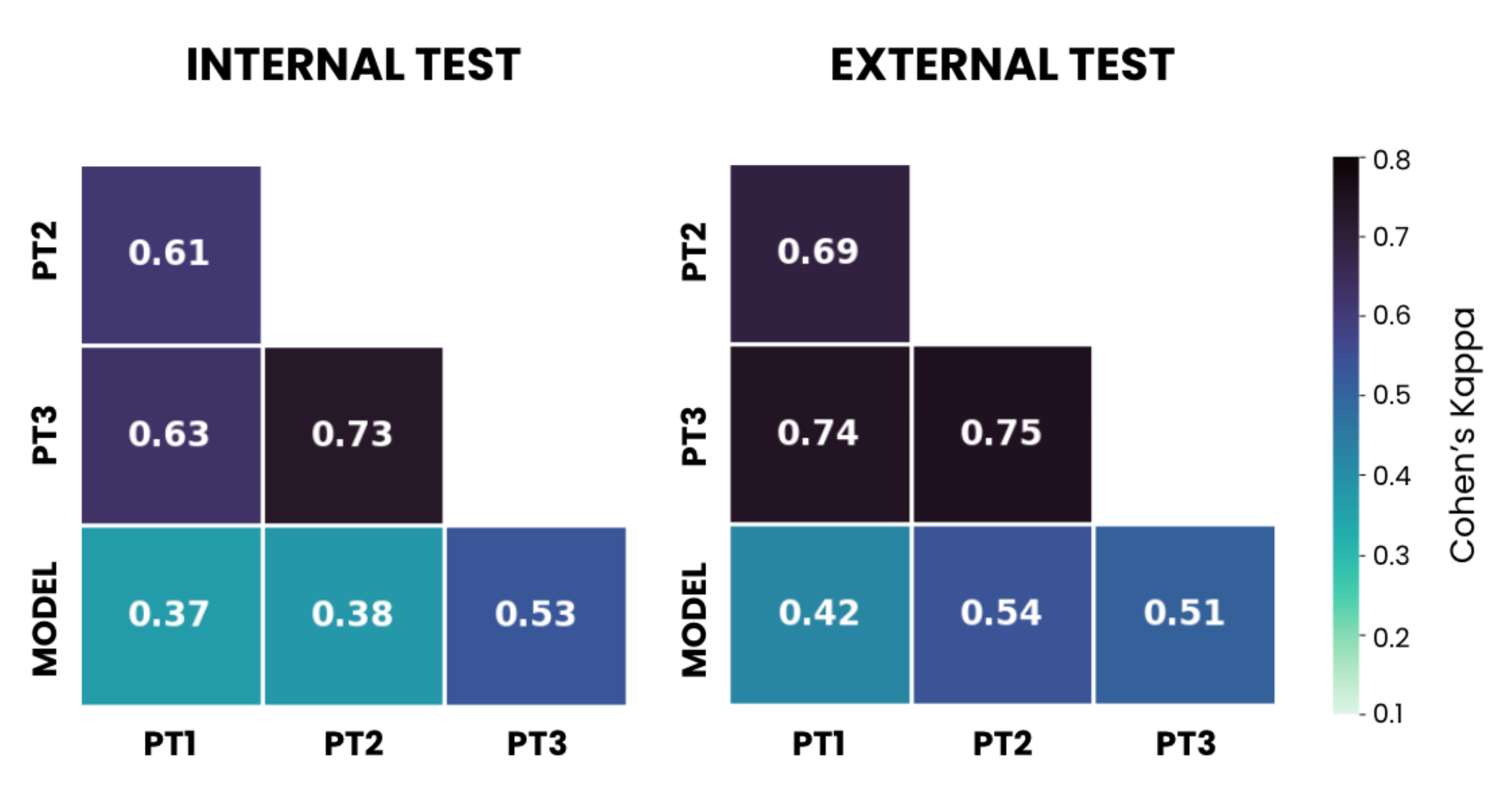}
    \caption{Inter-observer agreement in the assignment of OLGIM stages, compared with the automatic staging on the internal and external test datasets.}
    \label{fig:olgimagreement}
\end{figure}

Subplots in Figure~\ref{fig:olgimagreement} show the inter-observer agreement among pathologists, as well as between the pathologists and the best-performing deep learning model, regarding the OLGIM staging. Inter-observer agreement can be seen to increase when the categorical evaluation scale is simplified. In this work, expert agreement improved significantly when using a coarser grading scheme, the OLGIM staging system.

\subsubsection*{Variability of intestinal metaplasia percentages}
\label{subsec:resuncertainty}

\begin{figure}[htp]
    \centering
    \includegraphics[width=1\linewidth]{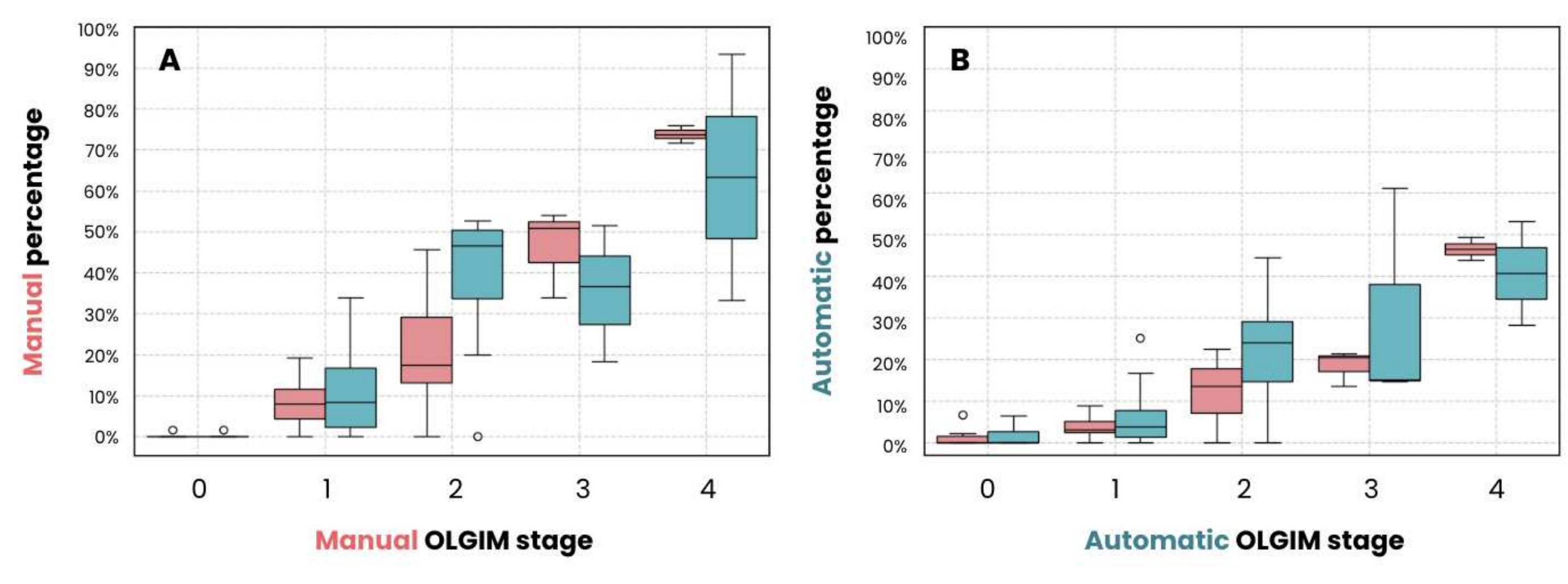}
    \caption{Distribution of manual (left) and automatic (right) intestinal metaplasia percentages in the antrum (red) and corpus (green), related to the consensus OLGIM stage assigned by pathologists.}
    \label{fig:distributions}
\end{figure}

Expert estimations of IM percentages showed the greatest variability in intermediate OLGIM stages, likely due to the larger number of possible antrum–corpus combinations that can yield the same stage. Both manual and automated assessments demonstrated a consistent increase in IM percentages with higher OLGIM stages. However, the automated model provided substantially lower variability across the four stages, while pathologists tended to assign higher percentages, reflecting a tendency to overestimate lesion extent. These differences are illustrated in Figure~\ref{fig:distributions}.

\section*{Discussion}
\label{sec:discussion}

This paper has investigated the role of state-of-the-art deep learning models to detect and quantify intestinal metaplasia (IM) on digitized biopsy samples from patients and volunteers. Evaluation of these models demonstrated relatively accurate performance for computing stroma, glands, and the percentage of IM, with F1 = $0.80$ and AUC = $0.91$. Comparison of OLGIM scoring, pathologist assessments, and deep learning predictions highlighted the inherent subjectivity of visual estimation, with only moderate inter-observer agreement (Fleiss’ Kappa = $0.31$ for the antrum and $0.41$ for the corpus), consistent with previously reported values in the literature~\cite{Lerch2022} (Fleiss’ Kappa = $0.48$). These findings confirm both the limitations of current risk systems, limiting the certainty of the medical decision and patient confidence, and the potential of AI tools to complement pathologist-based evaluation by serving as reproducible “second readers”.

Several studies have investigated OLGIM limitations, and many pathologists worldwide hardly report OLGIM stages since they are aware of the conflict introduced by the typical disagreement~\cite{Fang2024}, indeed, a moderate agreement is widely acknowledged~\cite{Salazar2022,Lerch2022}. An inevitable criticism of the OLGIM score is that it is inherently biased, that is, intermediate stages are obtained with a higher number of combinations and therefore these stages are more frequent, in other words “dices are charged” and confidence therefore is undermined. Additionally, pathologists tended to overestimate the extent of the lesion~\cite{Yue2018}, typically assigning higher percentages than the model. In the present investigation, the analysis was restricted to cross-sectional biopsies, limiting thereby the impact of the present study. Moreover, annotations represented a bottleneck, since delineated regions often included not only metaplasia but also stroma and normal glands, introducing bias~\cite{Selvaraju2017}. The multifocal nature of IM and its patch-like distribution further complicates assessment, as even standardized sampling with the five-biopsy protocol may fail to capture the full extent of lesions, affecting both manual and automated quantification. This work did not distinguish between complete and incomplete IM subtypes, which are morphologically and prognostically distinct, due to the lack of annotations at this level of detail. Finally, the participating pathologists were recently graduated, which may partly explain the relatively low inter-observer agreement observed.

It is important to recall that, in this work, OLGIM staging relied exclusively on IM since atrophy was difficult to evaluate, and even among experienced pathologists it has shown poor inter-observer reproducibility~\cite{Isajevs2014}. In contrast, IM has been recognized as the most reliable and prognostic component of OLGIM staging~\cite{Rugge2011}. This methodological choice therefore prioritizes objectivity and reproducibility. Recent studies have further demonstrated the effectiveness of deep learning models at classifying and grading pathologies in H\&E images~\cite{Coudray2018,LiuM2022,Wang2021}. Typically, pathologist knowledge, “the ground truth”\cite{Janowczyk2016}, consists of delineated regions that train models to replicate visual annotations. However, variable biology along with subjective judgment inevitably introduce additional biases.

Despite the multifocal nature of IM, i.e. spreading along stomach regions, estimation of this condition is often reduced to a binary problem: presence or absence. Both IM extension and topographic distribution have been at the very base of risk estimation for progression. Correa et al. highlighted that IM comprises two main subtypes, complete and incomplete, associated with distinct morphological expressions of mucin enzymes. Complete IM resembles the small intestine, with enterocytes displaying a brush border, whereas incomplete IM is closer to colonic epithelium with irregular intra-cytoplasmic mucin droplets of variable size. In the present study, we did not differentiate between these subtypes because annotations at this level of detail were not available, and our analysis focused on overall IM quantification in line with OLGIM staging. We acknowledge this as a limitation, and future research with dedicated annotations and tailored AI methods may enable reliable distinction of IM subtypes, moving beyond binary characterization and towards a more continuous description of the phenomenon, which is a necessary step for reproducible quantification and personalized risk estimation.

In summary, deep learning-based quantification of IM demonstrates robust performance, reproducibility, and potential clinical relevance. While expert variability, annotation bias, and sampling constraints remain significant challenges, computational pathology offers a pathway for systematic and objective evaluation. With further validation in multi-center cohorts, follow-up studies, and the integration of IM subtyping, AI-based tools could transform gastric cancer prevention strategies, moving from subjective visual estimation to reproducible, data-driven, and patient-tailored risk assessment.

\section*{Conclusion}
\label{sec:conclusion}

Deep learning models can reliably detect and quantify intestinal metaplasia with high performance, offering consistent estimates that may reduce the subjectivity of expert-dependent visual assessment. Further validation in larger, multi-center cohorts and the inclusion of complete and incomplete subtypes will be essential to capture the full morphological spectrum. In the long term, reproducible AI-based quantification of precancerous lesions has the potential to enhance gastric cancer risk stratification and guide more objective, personalized clinical decisions.

\subsubsection*{Author contributions}
\label{par:contributions}

Conceptualization: FC, ACR, FAG, SEV and ER; Data curation: FC, MC, AS, JV and JQ; Methodology: FC, MC, ACR, FAG and ER; Formal analysis: FC, ACR, FAG, SEV and ER; Writing - original draft preparation: FC and ER; Writing - review and editing: FC, ACR, FAG, SEV and ER; Funding acquisition: ACR and ER; Resources: ABU, MCB, YYC, ACR, FAG and ER; Supervision: FAG, SEV and ER; Project administration: ER. All authors have read and agreed to the published version of the manuscript.

\subsubsection*{Funding}
\label{par:funding}

This work was partially supported by the project with code 110192092345 \textit{Program for the Early Detection of Premalignant Lesions and Gastric Cancer in urban, rural and dispersed areas in the Department of Nariño} of call No. 920 of 2022 of MinCiencias. This work was partially supported by project 52895, titled \textit{Proposal for the strategic plan for the establishment of the Center of Excellence (Inter-Sites) in Medicine and Artificial Intelligence (SemAI)}, from the National Call for Proposals Bank for the Consolidation of Centers of Excellence 2020-2021 at \textit{Universidad Nacional de Colombia}. This work was partially supported by project BPIN 2019000100060 \textit{Implementation of a Network for Research, Technological Development and Innovation in Digital Pathology (RedPat) supported by Industry 4.0 technologies} from FCTeI of SGR resources, which was approved by OCAD of FCTeI and MinCiencias.

\subsubsection*{Acknowledgements}
\label{par:acknowledgements}

Special thanks to the CIEDYN foundation and the project BPIN 20150000100064 Urkunina5000, from which whole slides of gastric tissue were recovered from asymptomatic volunteers and subsequently digitized to be used in the development of this work.

\subsubsection*{Data availability}

The collection of internal gastric samples used in this study were obtained from the primary research project entitled \textit{Investigación de la prevalencia de lesiones precursoras de malignidad gástrica y efecto de la erradicación de Helicobacter pylori colon prevención primaria de cáncer gástrico en el Departamento de Nariño}. This project was approved under Agreement No. 057 of 2017 by the Collegiate Body for Administration and Decision (OCAD Pacífico), and was funded by the Science, Technology, and Innovation Fund of the General System of Royalties/Government of Nariño. Internal and external data are publicly available at Harvard Dataverse \cite{database}.

\section*{Declarations}

\subsubsection*{Conflict of interest}
\label{par:conflict}

All authors declare no conflicts of interest.

\subsubsection*{Ethical considerations}
\label{par:ethics}

This research study was conducted retrospectively using data provided by the CIEDYN foundation, a partner of the Urkunina5000 project, which contains information on ethical considerations in compliance with the Declaration of Helsinki. All patients signed informed consent forms. Additional ethics considerations was not required.

\subsubsection*{Informed consent}
\label{par:consent}

All asymptomatic volunteers selected for this study signed informed consent and all guarantees of anonymization were applied to their data.

\bibliography{manuscript}

\end{document}